\documentclass{mn2e}
\usepackage{times}
\usepackage{epsfig}

\newcommand{\beq}{\begin{equation}}
\newcommand{\eeq}{\end{equation}}
\newcommand{\lab}{\label}

\newcommand{\1}{\Omega_{\rm M0}}
\newcommand{\2}{\Omega_{\rm M}}

\begin{document}

\title{Constraints on a quintessence model from gravitational lensing statistics}

\author[M. Sereno]
{M. Sereno$^{1,2,3}$\thanks{E-mail: Mauro.Sereno@na.infn.it}
\\
$^{1}$Istituto Nazionale di Astrofisica - Osservatorio Astronomico di
Capodimonte, Salita Moiariello, 16, 80131 Napoli, Italia
\\
$^{2}$Dipartimento di Scienze Fisiche, Universit\`{a} degli Studi di
Napoli ``Federico II", Via Cinthia, Monte S. Angelo, 80126 Napoli,
Italia
\\
$^{3}$Istituto Nazionale di Fisica Nucleare, Sez. Napoli, Via Cinthia,
Monte S. Angelo, 80126 Napoli, Italia}

\maketitle

\begin{abstract}
Constraints on an exact quintessence scalar-field model with an
exponential potential are derived from gravitational lens statistics.
An exponential potential can account for data from both optical quasar
surveys and radio selected sources. Based on the Cosmic Lens All-Sky
Survey (CLASS) sample, lensing statistics provides, for the
pressureless matter density parameter, an estimate of $\1 =
0.31_{-0.14}^{+0.12}$.

\end{abstract}

\begin{keywords}
gravitational lensing -- cosmology: theory -- cosmological parameters
-- dark matter -- quasars: general
\end{keywords}

\section{Introduction}
Recently, astronomers have suggested that the universe is roughly
geometrically flat and accelerating its expansion. Usual types of
matter, i.e. baryons, photons and neutrinos, fail to close the
universe and generate attraction, which leads to a decelerated
expansion. To reconcile observations with predictions of general
relativity, two new components have been added to the energy budget of
the universe: pressureless cold dark matter (CDM) and dark energy,
also known as quintessence, with negative pressure.

Whereas the simplest explanation for dark energy is a cosmological
constant, one of the more interesting physical realizations of the
quintessence is a cosmic scalar field minimally coupled with the usual
matter action \cite{pe+ra88,cal+al98}. Such a field induces the
repulsive gravitational force dynamically, explaing the accelerated
expansion of our universe. $k$-essence, namely a scalar field with non
canonical kinetic terms \cite{arm+al99}, models based on branes and
extra dimensions, such as the Cardassian scenario \cite{zh+fu04}, and
the generalized Chaplygin gas \cite{kam+al01,zhu04} can also drive an
accelerated expansion.

A flat model of universe with a subcritical CDM energy density $\1$,
with two-thirds of the critical density existing in the form of dark
energy and with negligible amount in standard matter, matches
observations from galaxy clustering \cite{ba+fa98,car&al98},
large-scale structure \cite{pe&al01,ver+al01}, cosmic microwave
background radiation \cite{deb&al00,bal+al00,ja&al00,pr+al02}, type Ia
supernovae \cite{ri&al98,pe&al99}, age of the universe and Ly$\alpha$
forest \cite{ha&ro01,wan+al00}. So, to explain the bulk of the
evidence, we require a universe filled, nearly exclusively, with
exotic matter.

Gravitational lensing, based on the quite simple and well understood
physics of light deflection in a weak gravitational field, provides
useful tools for constraining cosmological parameters. Gravitational
lensing statistics
\cite{cha+al02,wa+mi98,co&hu99,wan+al00,zhu00,jai+al01}, effects of
large-scale structure growth in weak lensing surveys \cite{be+be01},
Einstein rings in galaxy-quasar systems \cite{fu&yo00,ya+fu01},
clusters of galaxies acting as lenses on background high redshift
galaxies \cite{io02cl,se+lo04}, offer very promising opportunities to
probe quintessence. Gravitational lens time delay measurements can
also produce precise estimates of cosmological parameters
\cite{sch04}, and the effect of quintessence in such observations has
been investigated \cite{gi+am01,le+ib02}. Results from techniques
based on gravitational lensing  are complementary to other methods and
can provide restrictive limits on the dark energy contribution,
sometimes in disagreement with the concordance value \cite{koc96}.

In this paper, gravitational lens statistics is used to study a class
of quintessence models with an exponential potential. Such a potential
can drive a scaling solution at late time, i.e. the scalar field
mimics the scaling of the dominant energy component. Usually,
associated with an exponential potential, a scalar field is considered
such that its fractional density, $\Omega_\varphi$, is practically
constant during part of the matter-dominated era. The usual view is
that, $\Omega_\varphi$ being small at the beginning of the
matter-dominated era due to bounds from nucleosynthesis theory, the
exponential potential cannot produce a today accelerated expanding
universe \cite{bea+al01}. This conclusion is mainly based on
considering the equation of state of the dark energy as nearly
constant. Recently, Rubano \& Scudellaro \shortcite{ru+sc02} showed
how, without such an a priori assumption, the exponential potential
can reveal very interesting features. This new position in favour of
the exponential potential relies on a very strong argument, i.e. on
general exact solutions of the field equations, allowed by a suitable
choice of the exponent of the scalar potential. It has been shown
that, since the equation of state is not constant, the tracker
condition cannot be treated in the usual way. Despite of its
simplicity, this class of potentials is able to fit data from type Ia
supernovae \cite{pav+al02}, galaxy clustering \cite{ru+se02} and
cosmic microwave background \cite{did+al02}.

The paper is as follows. In Section~\ref{mode}, the quintessence model
is introduced. Section~\ref{stat} contains the basics of gravitational
lensing statistics. Properties of the data sample used in the
statistical analysis are illustrated in Sec.~\ref{data}.
Section~\ref{resu} lists the constraints on the cosmological
parameters. Section~\ref{disc} is devoted to some final
considerations.

\section{Model description}
\label{mode}

A spatially flat, homogeneous and isotropic universe filled with two
non-interacting components, pressureless matter (dust) and a scalar
field $\varphi$ minimally coupled with gravity, can account for the
present day observational data. Here, we investigate a quintessence
model with a potential of the form \cite{ru+sc02}
\beq
\lab{scal1}
V(\varphi) \propto \exp\left\{ -\lambda \varphi\right\},
\eeq
where $\lambda^2 \equiv 12 \pi G/ c^2$. The exponential potential has
been widely studied, in relation with both quintessence and inflation
\cite{pe+ra88,cap+al96}. The particular choice of $\lambda$ allows for
general exact integration of the cosmological equations. I present
only what is needed for the present work. The time dependent Hubble
parameter $H$ is
\beq
\lab{scal2}
H=\frac{2(1+2\tilde{t}^2)}{3t_{\rm s} \tilde{t} (1+\tilde{t}^2)},
\eeq
and the pressureless matter density ($\rho_{\rm M}$), in units of the
critical density $\rho_{{\rm cr}} \equiv 3H^2 /8\pi G$, is
\beq
\lab{scal3}
\2 =\frac{1+\tilde{t}^2}{(1+2\tilde{t}^2)^2},
\eeq
where $\tilde{t} \equiv t/t_{\rm s}$ is a dimensionless time and
$t_{\rm s}$ is a time scale of the order of the age of the universe.

The relation between the dimensionless time $\tilde{t}$ and the
redshift $z$ is given by
\beq
\lab{scal4}
(1+z)^3=\frac{\tilde{t}_0^2 (1+\tilde{t}_0^2)}{\tilde{t}^2
(1+\tilde{t}^2)},
\eeq
where $\tilde{t}_0$ is the present value of $\tilde{t}$. This very
simple cosmological model has two free parameters, $t_{\rm s}$ and
$\tilde{t}_0$, or, equivalently, $H_0$, the present value of the
Hubble parameter, and $\1$, the present value of the pressureless
matter density. As can be seen from Eq.~(\ref{scal3}), $\1$ depends
only on $\tilde{t}_0$.

The angular diameter distance between $z_{\rm d}$ and $z_{\rm s}$ in a
flat universe is, in units of $c/H_0$,
\beq
\label{dist}
d_{\rm A}(z_{\rm d},z_{\rm s})=\frac{1}{1+z_{\rm s}}
\int_{z_{\rm d}}^{z_{\rm s}}
\frac{H_0}{H(z)}dz,
\eeq
where, now, $H$ denotes the Hubble parameter as a function of the
redshift. The dimensionless angular diameter distance in
Eq.~(\ref{dist}) depends only on $\1$. For cosmological distances in
an inhomogeneous quintessence cosmology, I refer to Sereno et
al.~\shortcite{ser+al01,ser+al02}

\section{Lensing statistics}
\label{stat}

To calculate the statistics of gravitational lenses, simple
assumptions are made \cite{koc93,koc96}. The standard approach is
based on observed number counts of galaxies, on some empirical
relations between velocity dispersion and absolute magnitude, and,
finally, on the simple singular isothermal sphere (SIS) model for lens
galaxies.

The differential probability of a beam of intersecting a galaxy with
velocity dispersion between $\sigma$ and $\sigma + d\sigma$, at
redshift $z_{\rm d}$ in the interval $d z_{\rm d}$ is
\beq
\frac{d^2 \tau}{d z_{\rm d} d \sigma} = \frac{d n_{\rm G}}{d \sigma}(z_{\rm d},\sigma) s_{\rm
cr}(\sigma)
\frac{cd t}{d z_{\rm d}} ,
\eeq
where $s_{\rm cr}$ is the cross section for lensing event, and
$\frac{d n_{\rm G}}{d \sigma}$ is the differential number density
population. For a conserved comoving number density of lenses, $n_{\rm
G}(z) =n_0(1+z)^3$.

As usual in lensing statistics, all lenses are associated with
optically luminous galaxies. The luminosity function (LF) of galaxies
can be modeled by a Schechter function of the form \cite{sch76}
\beq
\frac{dn}{d(L/L_*)}=n_*\left( \frac{L}{L_*}\right)^\alpha \exp \left[ -\frac{L}{L_*}\right],
\eeq
where $\alpha$ is the faint-end slope and $n_*$ and $L_*$ are the
characteristic number density and luminosity, respectively. The
parameter $\alpha$ characterizes the faint part of the LF, which is
still uncertain; a value $\sim -1$ implies the existence of numerous
faint galaxies acting as lenses.

Since early-type and late-type populations contribute to the multiple
imaging in different ways, type-specific LFs are required. As a
conservative approach, I do not consider lensing by spiral galaxies,
as their velocity dispersion is small in comparison to E/S0 galaxies.
The luminosity of a galaxy is correlated with its line-of-sight
stellar velocity dispersion $\sigma$ via the empirical relation
\beq
\frac{L}{L_*}=\left( \frac{\sigma}{\sigma_*}\right)^\gamma ,
\eeq
where $\sigma_*$ is the characteristic velocity dispersion. The
exponent $\gamma$, for early-type galaxies, can be fixed to the
Faber-Jackson value, $\gamma =4$ \cite{fa+ja76}.

Early-type galaxies are well approximated as singular isothermal
spheres. As shown in Maox \& Rix \shortcite{ma+ri93} and Kochanek
\shortcite{koc96}, radial mass distribution, ellipticity and
core radius of the lens galaxy are unimportant in altering the
cosmological limits. Assuming a flat model of universe, a typical
axial ratio of 0.5 in a mixed population of oblate and prolate
spheroids would induce a shift of $\sim 0.04$ in the estimation of
$\1$ \cite{mit+al04}, well below statistical uncertainties. Since
departures from spherical symmetry induce a relatively small effect on
lens statistics and the distribution of mass ellipticities is highly
uncertain, spherically symmetric models supply a viable approximation.
The cross section of a SIS is
\beq
s_{\rm cr}=4\pi^2 \left( \frac{\sigma}{c}\right)^4  \left(
\frac{c}{H_0} \right)^2 \left( \frac{D_{\rm d} D_{\rm ds}}{D_{\rm s}} \right)^2,
\eeq
where $D_{\rm d}$, $D_{\rm ds}$ and $D_{\rm s}$ are the angular
diameter distances between the observer and the deflector, the
deflector and the source and the observer and the source,
respectively.

When the deflector population is modeled as SISs and it is distributed
according to the Schechter function, the differential probability for
a source at $z_{\rm s}$ to be multiple imaged with image separation
$\Delta \theta$ by a galaxies between $z_{\rm d}$ and $z_{\rm d}
+dz_{\rm d}$ is
\begin{eqnarray}
\frac{d^2 \tau}{d z_{\rm d} d \Delta \theta} & = & 8 \pi^3 n_* \gamma \left( \frac{\sigma_*}{c}
\right)^4
\left( \frac{c}{H_0}\right)^3 (1+z_{\rm d})^2\frac{H_0}{H(z_{\rm d})} \nonumber \\
& {\times} &
\left( \frac{D_{\rm d} D_{\rm ds}}{D_{\rm s}} \right)^2
\left( \frac{D_{\rm s}}{D_{\rm ds}} \hat{\Delta \theta}
\right)^{\frac{\gamma}{2}(1+\alpha)+1} \nonumber \\
& {\times} & \exp \left[
\left( -\frac{D_{\rm s}}{D_{\rm ds}} \hat{\Delta \theta} \right)^\frac{\gamma}{2}\right]
\frac{1}{\Delta \theta_* \frac{D_{\rm s}}{D_{\rm ds}}},
\end{eqnarray}
where $\Delta \theta_* \equiv 8 \pi \left(
\frac{\sigma_*}{c}\right)^2$ and $\hat{\Delta \theta} \equiv \frac{\Delta \theta}{ \Delta
\theta_*}$.

The total optical depth for multiple imaging a compact source, in a
flat universe, is
\beq
\tau(z_{\rm s})=\frac{F_*}{30}\left[ (1+z_{\rm s}) d_A(0,z_{\rm s})\right]^3,
\eeq
where $F_* \equiv 16 \pi^3 n_* \left( \frac{c}{H_0}\right)^3
\left( \frac{\sigma_*}{c}\right)^4 \Gamma [1+\alpha +4/\gamma]$.

The configuration probability that a SIS in a flat universe forms
multiple images of a background source with angular separation $\Delta
\theta$ is \cite{koc93}
\begin{eqnarray}
p_{\rm c} (\Delta \theta) d \Delta \theta & = &
\frac{1}{\tau}\int_0^{z_{\rm s}}
\frac{d^2 \tau}{d z_{\rm d} d \Delta \theta} d z_{\rm d} \\
&  = & 30 \frac{\hat{\Delta
\theta}^2}{ \Delta \theta_* }  \frac{d\Delta \theta}{\Gamma [1+\alpha+4\gamma^{-1}]}  \\
& {\times} &
\left\{
 \Gamma \left[ 1+\alpha -2\gamma^{-1} , \hat{\Delta \theta}^\frac{\gamma}{2}\right] \right.  \nonumber \\
 & -&  2\hat{\Delta \theta}\
\Gamma \left[ 1+\alpha -4\gamma^{-1} , \hat{\Delta \theta}^\frac{\gamma}{2}\right] \nonumber \\
 & + & \left. \hat{\Delta \theta}^2\
 \Gamma \left[ 1+\alpha -6\gamma^{-1} , \hat{\Delta \theta}^\frac{\gamma}{2}\right]
\right\}. \nonumber
\end{eqnarray}

Two corrections to the optical depth must be included: magnification
bias and selection function. Magnification bias accounts for tendency
of gravitationally lensed sources to be preferentially included in
flux-limited samples due to their increased apparent brightness
\cite{tur90,fu+tu91,fuk+al92,koc93}. The bias factor for a source at
redshift $z_{\rm s}$ with apparent magnitude $m$ is given by
\begin{eqnarray}
{\bf B}(m,z, M_0) & =&  \left(
\frac{dN_{\rm s}}{dm}\right)^{-1} \\
& {\times} & \int_{M_0}^{+\infty}
\frac{dN_{\rm s}}{dm}(m+2.5\log M,z)P(M)dM , \nonumber
\end{eqnarray}
$M_0$ being the minimum magnification of a multiply imaged source,
with value $M_0 =2$; $P(M)dM =2 M_0^2M^{-3}dM$ is the probability that
a multiple image-lensing event causes a total flux increase by a
factor $M$ \cite{koc93}. The function $dN_{\rm s}/d m$ is the
differential source number count in magnitude bins $dm$. The
magnification corrected probabilities are
\beq
p(m,z_{\rm s}) =\tau(z_{\rm s}) {\bf B}(m,z_{\rm s}).
\eeq

Lens discovery rates are affected by the ability to resolve multiple
source images \cite{koc93}. Since observations have finite resolution
and dynamic range, a selection function must be included to correct
the statistics for observational limitations. For the SIS model,
selection effects can be characterized by the maximum magnitude
difference that can be detected for two images separated by $\Delta
\theta$, $\Delta m(\Delta \theta)$, which determines a minimum total
magnification $M_{\rm f} =M_0(f+1)/(f-1)$, where $2.5 \log f \equiv
\Delta m$ \cite{koc93}. Then, the corrected lensing probability is \cite{koc93,koc96}
\begin{eqnarray}
p^{'}(m,z)& = & p(m,z)\int\frac{ {\bf B}(m,z,M_{\rm f}(\Delta
\theta))}{{\bf B}(m,z,M_0)}p_{\rm c}(\Delta \theta) d\Delta \theta \nonumber \\
& = & \tau(z) \int {\bf B}(m,z,M_{\rm f}(\Delta
\theta)) p_{\rm c}(\Delta \theta) d\Delta \theta
\end{eqnarray}
The sum of the lensing probabilities $p^{'}$ for the selected sample
gives the expected number of lensed sources,
\beq
\label{num}
n_{\rm L} =\sum_i p^{'}_i
\eeq
where $p^{'}(m_i,z_i) \equiv p^{'}_i$.

The corrected image separation distribution function, $p_{\rm c}^{'}$,
describes the probability that a source is lensed with the observed
image separation. For a single source at $z$, it is \cite{koc93,koc96}
\begin{eqnarray}
p_{\rm c}^{'} (\Delta \theta, m,z)& = & p_{\rm c} (\Delta
\theta)\frac{p(m,z)}{p^{'}(m,z)} \frac{ {\bf B}(m,z,M_{\rm f}(\Delta
\theta))}{{\bf B}(m,z,M_0)} \nonumber \\
& = & p_{\rm c} (\Delta \theta) \frac{ {\bf B}(m,z,M_{\rm f}(\Delta
\theta))}{ \int {\bf B}(m,z,M_{\rm f}(\Delta
\theta)) p_{\rm c}(\Delta \theta) d\Delta \theta }
\end{eqnarray}

I can now introduce the likelihood function \cite{koc93,cha+al02},
\beq
{\cal L}= \prod_{i=1}^{N_{\rm U}}(1-p^{'}_i)\prod_{j=1}^{N_{\rm
L}}p_{l,j},
\eeq
where $N_{\rm L}$ is the number of multiple-imaged sources and $N_{\rm
U}$ is the number of unlensed sources. $p_l$ is the suitable
differential probability accounting for the whole of the data
available for each lens system, i.e. the lens redshift and/or the
image separation \cite{cha+al02,mit+al04}. I use a uniform
distribution for the prior on the cosmological parameter $\1$, so
that, apart from an overall normalization factor, the likelihood can
be identified with the posterior probability.

\section{Data sample}
\label{data}

Reliable cosmological constraints from gravitational lens statistics
need an unbiased sample. Indeed, a statistical sample must be complete
and selected with well-defined criteria. The statistical properties of
gravitational lensing strongly depend on the properties of the
distributions of sources in redshift and in luminosity in the
observational selection wave band.

Analyses of statistical lensing have been based either on optically
selected samples or radio-selected sources. The two methods present
both advantages and problems. Radio surveys are successful in
selecting homogeneous and complete samples of sources. Since a radio
source is selected purely on its radio flux and spectral index and the
radio properties of a lens are unaffected by optical properties of the
lens galaxy or extinction in the lens galaxy, radio surveys can avoid
some systematic errors that can affect results based on quasar lens
surveys at optical wavelengths. The main shortcoming with
radio-selected samples is the lack of detailed information on the
global properties of their redshift and luminosity distribution. On
the other hand, optical samples require problematic matching of
independent data sets \cite{koc96} and suffer problems concerning
extinction, but have a full descriptions of the properties of
individual sources. In order to take advantages of the properties of
observations in the two wave-bands, in this paper the statistical
analysis will be performed with both the optical- and radio-selected
samples.

A proper modeling of the distribution of the lensing galaxies is
central in lensing statistics. Different points of view have been
expressed on the best choice for the distribution in the deflector
population \cite{mit+al04,cha+al04}. However, results from several
recent galaxy surveys containing a large number of galaxies appear to
converge towards concordant estimates \cite{cha03}. Chae
\shortcite{cha03} used data from the Second Southern Sky Redshift
Survey (SSRS2) to derive relative LFs. The early-type LF, expressed in
the $B$ photometric system, has: $\alpha=-1.00 \pm 0.09$, $n_*
=(6.4 \pm 1.9){\times} 10^{-3}h^3$~Mpc$^{-3}$, where $h$ is $H_0$ in units of
100~km~s$^{-1}$~Mpc$^{-1}$. The early-type characteristic
velocity dispersion can be fixed at $\sigma_*=192 {\pm}34$~km~s$^{-1}$
\cite{cha03}. In this work, the SSRS2 LF will be adopted.

\subsection{The Cosmic Lens All-Sky data sample}

\begin{figure}
        \epsfxsize=8cm
        \centerline{\epsffile{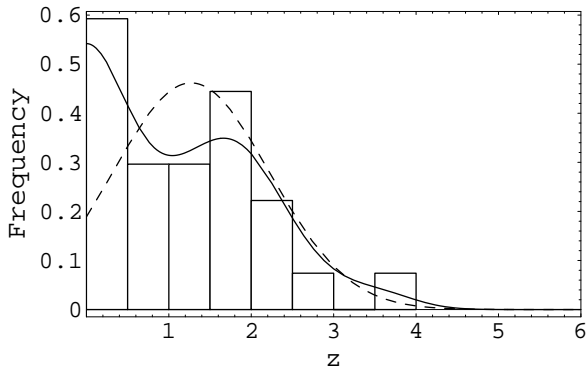}}
        \caption{Normalized redshift distribution of the flat spectrum sources in
        Marlow et al.~2001. The kernel estimator of the distribution
        (full line) and a Gaussian distibution with $\langle z_{\rm s}\rangle =1.27$ and
        dispersion 0.95 (dashed line) are overplotted.}
        \label{RedshiftSourcesCLASS}
\end{figure}

Radio-selected galactic mass-scale gravitational lens survey can
provide very large, homogeneous, unbiased statistical samples of
sources. The Cosmic Lens All-Sky Survey (CLASS, Browne et
al.~\shortcite{bro+al03,mye+al03}; Myers et al.~\shortcite{mye+al03})
is the largest radio selected galactic mass-scale gravitational lens
search project to date. Out of the about 16,000 imaged sources,
including the predecessor project Jodrell Bank Very Large Array
Astrometric Survey, a subsample of 8958 flat-spectrum radio sources
down to a 5~GHz flux density of 30~mJy, with 13 lenses constitutes a
well-defined subset suitable for statistical analysis \cite{bro+al03}.
I use the data listed in Table~1 of Chae~\shortcite{cha03} updated
with the spectroscopic observations by McKean et
al.~\shortcite{mck+al04}, which identified the lenses for 0445+123 and
0631+519 as early-type galaxies at $z_{\rm d}=0.558$ and 0.620,
respectively, while that for 0850+054 is a spiral-type galaxy at
$z_{\rm d}=0.588$.

I limit my analysis to the early-type lens galaxies, as in the
standard approach to lens statistics \cite{mit+al04}. Whereas the
description of the late-type galaxy population is plagued by large
uncertainties, they contribute no more than 10-20\% of the total
lensing optical depth. I exclude 0218+357 and 0850+054, whose lenses
have been certainly detected as spiral galaxies. There are also
arguments for discarding 1359+154 which presents a multiple lens
system with three deflecting galaxies \cite{mit+al04}. So, I perform
the statistical analysis considering two cases separately, the first
one with 11 lenses and the second one with 10 cases. I do not use the
measured image separations of 1359+154, 1608+656, and 2114+022, whose
splitting are due to multiple galaxy systems.

A proper lens statistics analysis requires an accurate knowledge of
the global properties of flat-spectrum radio sources. The final
statistical lens sample is well described at 5~GHz by a power-law
differential number-flux density, $\left| dN/d S_\nu
\right| \propto S_\nu^{-\eta}$, with $\eta = 2.07 {\pm} 0.02$ \cite{cha03}. When the
source counts can be modeled with a power law, the bias factor for a
SIS lens population reads
\beq
{\bf B}(m,z,M_{\rm f}) = \frac{2^\eta}{3-\eta}\left(
\frac{f-1}{f+1}\right)^{3-\eta},
\eeq
where $f$ is the maximum detected ratio of the flux densities of the
brighter to the fainter images. The final CLASS statistical sample has
been selected such that, for doubly imaged systems, the flux ratio is
$\leq 10$ and it is independent of the angular separation.

Redshift measurements are only available for a restricted CLASS
subsample. I model the redshift distribution of the unlensed sources
with a kernel empirical estimator \cite{vio+al94,ryd96}. Given a
sample of $N$ measured source redshifts, $\{ z_{{\rm s},i}
\}$,  the kernel estimator of the distribution is
\beq
\label{kern1}
N_z (z_{\rm s})=\frac{1}{N g} \sum_{i=1}^N K\left(
\frac{z_{\rm s}-z_{{\rm s},i}}{g}\right),
\eeq
where $K$ is the kernel function. Since the redshift is limited to non
negative values, I use the kernel
\beq
\label{kern2}
K_{\rm ref}(z_{\rm s},z_{{\rm s},i},g)= K_{\rm Gau}\left(
\frac{z_{\rm s}-z_{{\rm s},i}}{g}\right)+ K_{\rm Gau}\left(
\frac{z_{\rm s} + z_{{\rm s},i}}{g}\right),
\eeq
where $K_{\rm Gau}$ is a Gaussian kernel
\beq
K_{\rm Gau} (x)=\frac{1}{\sqrt{2 \pi}} e^{-x^2/2}.
\eeq
Using the kernel in Eq.~(\ref{kern2}), the Gaussian tail extending to
negative values of $z_{\rm s}$ is folded back into the positive axis.
The kernel width $g$ is well approximated by $g \simeq 0.9 A
N^{-0.2}$, with $A$ the smaller of the standard deviation of the
sample and the interquartile range divided by 1.34.

I consider the spectroscopic observations of a sample of 42
flat-spectrum radio sources from CLASS in Marlow et
al.~\shortcite{mar+al01}. The 27 sources with measured redshift have a
mean redshift of 1.27, with a standard deviation of $\sim 1$. In
Fig.~\ref{RedshiftSourcesCLASS}, the histogram of the source redshift
sample and the kernel estimator are shown together with a Gaussian
distribution also used to model the distribution \cite{cha03}. For the
unmeasured lensed source redshifts, I set $z_{\rm s}$ to the mean
source redshift for the lensed sources with measured redshift,
$\langle z_{\rm s} \rangle_{\rm lensed}=1.93$.

Finally, according to the CLASS selection criteria, the compact
radio-core images have separations gretar than $ \Delta
\theta_{\rm min}=0.3$~mas. The probabilities that enter the likelihood
must be considered as the probabilities of producing image systems
with separations $\geq\Delta \theta_{\rm min}$.

\subsection{The quasar optical sample}

As discussed above, radio lens surveys have little information on the
intrinsic redshift distribution of the sources. Since the optical
depth to multiple image scales as $D_{\rm s}^3$, this can lead to a
serious problem in statistical analyses of lens samples.
Kochanek~\shortcite{koc96b} showed the strong correlation between the
mean redshift of the fainter sources and the cosmological model, so
that uncertainties in the redshift distribution, or equivalently in
the radio luminosity function, lead to systematic uncertainties in the
derivation of cosmological limits. Furthermore, since the fraction of
the sources identifiable as quasars steadily drops as the flux density
of the sources decreases \cite{mun+al03}, any estimate of the radio
luminosity function needs to divide the sources into two populations,
quasars and galaxies, contrary to earlier studies. These shortcomings
will be addressed with systematic programs to estimate the redshift
distributions of the fainter flat-spectrum sources \cite{mun+al03}.

So, it still turns out to be very useful to perform statistical
analyses based on optical lens surveys. To perform the analysis, I use
data from six optical surveys for gravitationally lensed quasars: the
HST Snapshot survey \cite{mao+al93}, the HRCam survey \cite{cra+al92},
the Yee survey \cite{yee+al93}, the NOT survey \cite{jau+al95}, the
FKS survey \cite{koc+al95} and the ESO/Liege survey \cite{sur+al93}. A
total of 893 high luminous optical quasars, plus five lenses, is
considered \cite{koc96}. Matching different surveys requires a close
examination of the selection effects and reliability of the individual
data sample. I refer to Kochanek~\shortcite{koc93,koc96} for an
extensive discussion.

The differential quasar number count, $dN_{\rm Q}/dm$, can be modeled
as \cite{koc96}
\beq
\frac{dN_{\rm Q}}{dm} \propto \left( 10^{-a(m-m_0(z))} + 10^{-b(m-m_0(z))} \right)^{-1},
\eeq
where the bright-end slope is $a=1.07 \pm 0.07$ and the faint-end
slope is $b = 0.27 \pm 0.07$. The break magnitude $m_0(z)$ evolves as
\beq
m_0(z) = \left\{
\begin{array}{cc}
m_0+(z-1), & z \leq 1, \\ m_0, & 1 < z \leq 3, \\ m_0 -0.7(z-3), & 3 <
z,
\end{array}
\right.
\eeq
with $m_0 =18.92 \pm 0.16$ at $B$ magnitude. Since lens surveys and
quasar catalogues usually use $V$ magnitudes, I adopt an average $B-V$
colour of 0.2 \cite{bah+al92,mao+al93}. Finally, I use two selection
functions, as suggested in Kochanek \shortcite{koc93}, one for space
observations and another one for all the ground based surveys.

\section{Results}
\label{resu}

\begin{figure}
        \epsfxsize=8cm
        \centerline{\epsffile{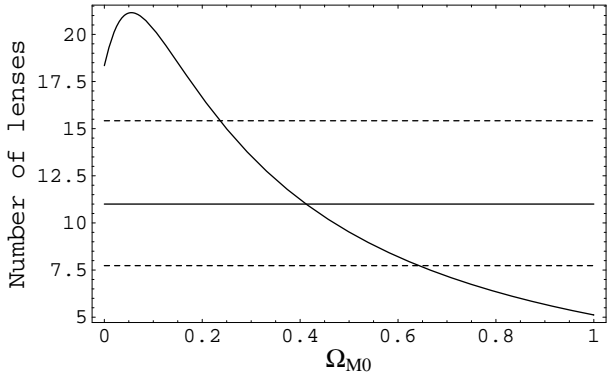}}
        \caption{Expected number of lenses as a function of $\1$ for the CLASS sample.
        Short-dashed horizontal lines are for the conventional 68.3\% confidence limit in the
        case 11 events (horizontal full line) are detected in a given observation.}
        \label{NumberOfLensesCLASS}
\end{figure}

\begin{figure}
        \epsfxsize=8cm
        \centerline{\epsffile{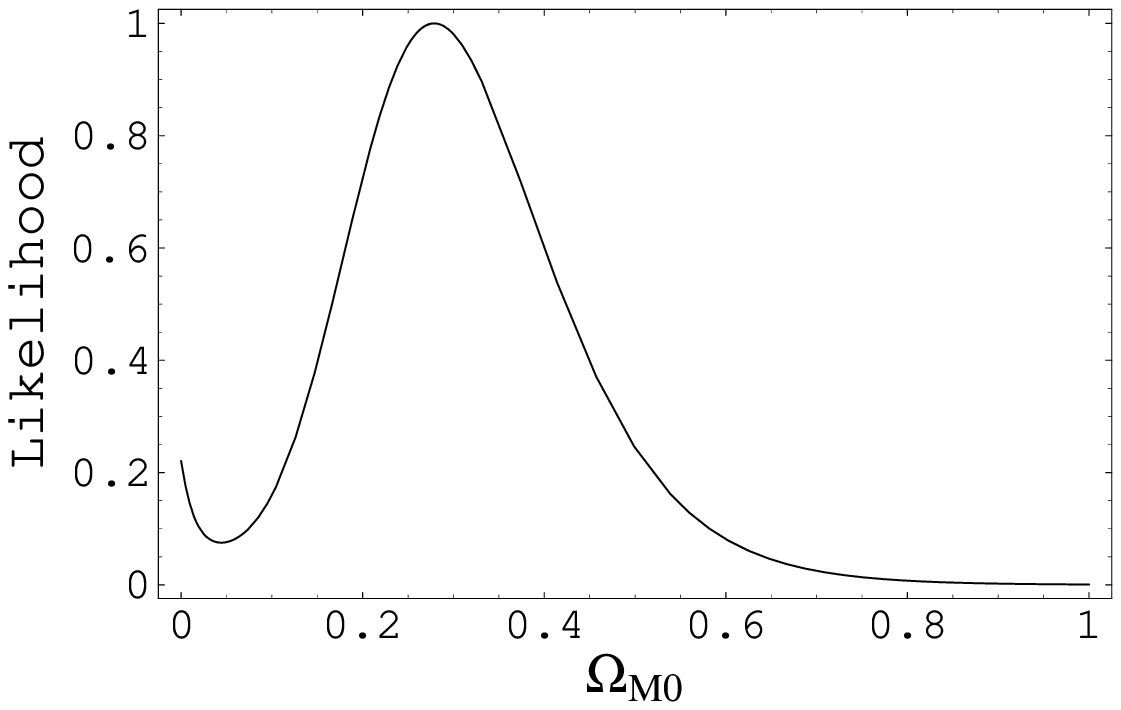}}
        \caption{Normalized likelihood function from lensing
        statistics as a function of $\1$ for the CLASS sample in the case
        of 11 lenses.}
        \label{LikelihoodCLASS11}
\end{figure}

\begin{figure}
        \epsfxsize=8cm
        \centerline{\epsffile{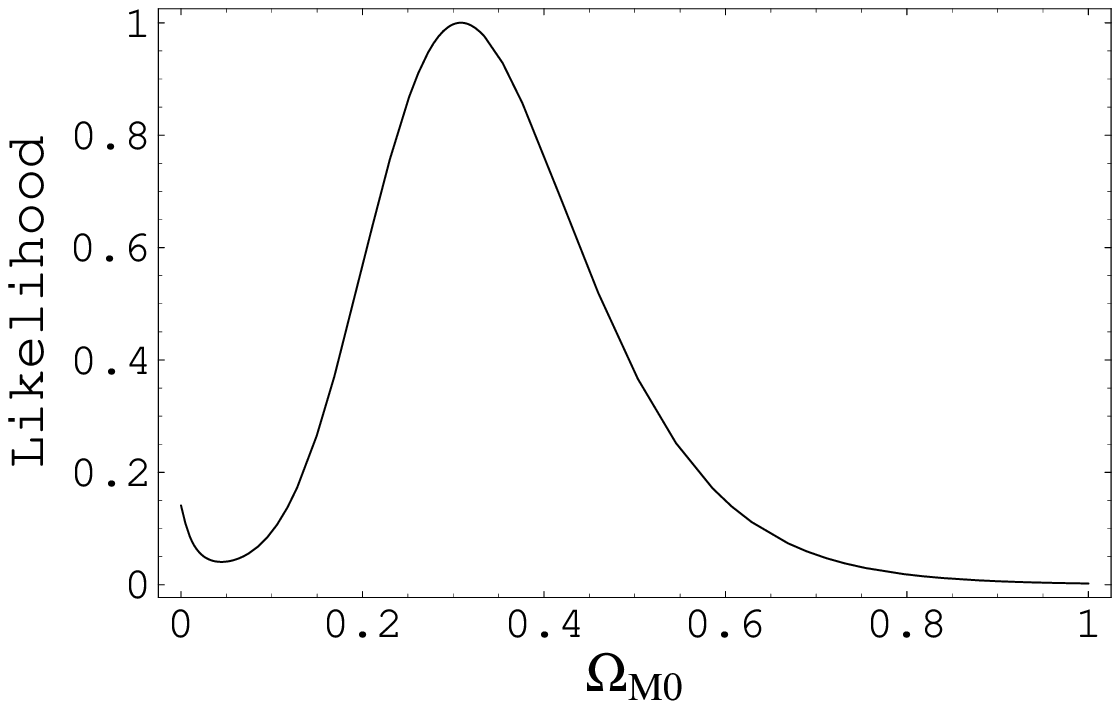}}
        \caption{Normalized likelihood function from lensing
        statistics as a function of $\1$ for the CLASS sample in the case
        of 10 lenses.}
        \label{LikelihoodCLASS10}
\end{figure}

\begin{figure}
        \epsfxsize=8cm
        \centerline{\epsffile{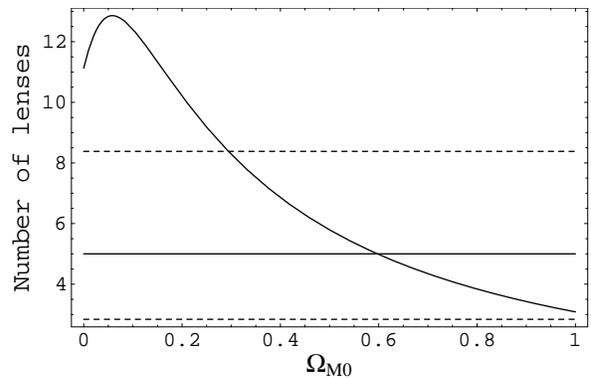}}
        \caption{Expected number of lenses as a function of $\1$ for the optical sample.
        Short-dashed horizontal lines indicate the conventional 68.3\% confidence limit in
        the case of 5 events (horizontal full line).}
        \label{NumberOfLensesQSO}
        \end{figure}

\begin{figure}
        \epsfxsize=8cm
        \centerline{\epsffile{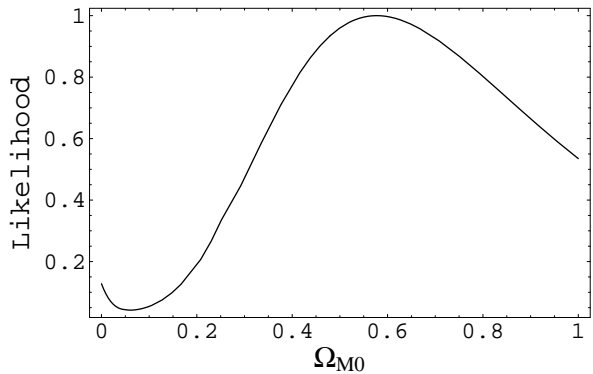}}
        \caption{Normalized likelihood function from lensing
        statistics as a function of $\1$ for the optical sample.}
        \label{LikelihoodQSO}
\end{figure}

Let us now perform a statistical analysis to determine the parameter
$\1$, which enter the model described in Section~\ref{mode}. We first
consider the radio sample and, then, the optical one.

\subsection{Radio sample}

Let us first evaluate the predicted number of lenses, Eq.~(\ref{num}).
In Fig.~\ref{NumberOfLensesCLASS}, I plot the predicted number of
lensed sources, $n_{\rm L}$, in the CLASS sample. The calculation of
confidence limits is based on standard equations derived from Poisson
statistics \cite{geh86}. In the case 11 events are detected in a given
observation, the conventional 68.3\% confidence limit is $ 7.734 \leq
n_s \leq 15.42$ \cite{geh86}. It is $n_{\rm L} = 11$ for $\1 =0.41$,
with $0.23
\stackrel{<}{\sim} \1 \stackrel{<}{\sim} 0.64$ at 68.3\% confidence limit.
In the case of 10 lenses, the 68.3\% confidence limit, $6.891 \leq
n_{\rm L} \leq 14.27$ \cite{geh86}, corresponds to $0.27
\stackrel{<}{\sim} \1
\stackrel{<}{\sim} 0.73$; $n_{\rm L} = 10$ for $\1 =0.47$.

As shown in Mitchell et al.~\shortcite{mit+al04}, the likelihood can
be approximated as
\beq
{\cal L} \simeq \exp \left[ -\int N_z(z_{\rm s})p^{'} (z_{\rm s})d
z_{\rm s}
\right]
\prod_{j=1}^{N_{\rm L}}p_{l,j}.
\eeq
The likelihood is maximized for $\1
= 0.28_{-0.13}^{+0.12}$ for 11 lenses, see
Fig.~\ref{LikelihoodCLASS11}, or $0.31_{-0.14}^{+0.12}$ for 10 lenses,
see Fig.~\ref{LikelihoodCLASS10}. Uncertainties denote the central
68.3\% interval around the mode. Since the likelihood functions are
slightly asymmetric, I also provide the expectation value and the
standard deviation of the distributions \cite{dag04}. I get $\1 = 0.31
{\pm}0.13$ for 11 lenses, or $0.35 {\pm} 0.14$ for 10 lenses.

The finite sample size induces an error in the estimated redshift
distribution. From the initial distribution estimated from the
original $N=27$ source redshifts, I created 1000 data sets each
containing $N$ points. Each data set is then used to create a new
kernel estimator for the redshift distribution.  The likelihood
analysis is then repeated using the new 1000 redshift distributions.
The resulting maximum likelihood estimates of $\1$ present a
dispersion of 0.08. I have also verified that, when modeling the
redshift distribution with a Gaussian distribution instead of the
kernel estimator \cite{cha03,mit+al04}, conclusions are really
unaffected.

The main uncertainty in the estimation of cosmological parameters
comes from uncertainties in the assumed parameters of the luminosity
function which describes the lens population. I created 1000 sets of
galactic parameters $n_*$, $\alpha_*$ and $\sigma_*$. The values of
the parameters are extracted from normal distributions centred on the
best estimates of each parameter and with standard deviation given by
the associated uncertainty. The likelihood analysis is then repeated
for each set of galactic parameters. The resulting distribution of
maximum likelihood estimates has a scatter of $\sim 0.3$, which give a
similar uncertainty in the overall determination of $\1$. The main
contribution comes from the uncertainty in the typical velocity
dispersion $\sigma_*$.

\subsection{Optical sample}

Let us first evaluate the predicted number of lenses, Eq.~(\ref{num}).
In Fig.~\ref{NumberOfLensesQSO}, I plot the predicted number of lensed
quasars in the adopted sample, $n_Q$. In the case 5 events are
detected in a given observation, the conventional 68.3\% confidence
limit is $ 2.480 \leq n_Q \leq 8.382$ \cite{geh86}. It is $n_Q = 5$
for $\1 =0.60$, with $\1 \stackrel{>}{\sim} 0.29$ at 68.3\% confidence
limit.

The expectation value and the standard deviation of the likelihood
distribution, see Fig.~\ref{LikelihoodQSO}, agree with the mode and
the central $68.3\%$ interval around the peak, respectively. It is $\1
= 0.6 {\pm} 0.2$. For $\1=0.6$, 5.1 lenses are predicted.

No clear upper bound can be derived from the analysis of the optical
sample. As can be seen from
Figs.~\ref{NumberOfLensesQSO},~\ref{LikelihoodQSO}, even the
Einstein-de Sitter model, where $\1=1$ and dark energy is not present,
is compatible with the data.

\section{Discussion}
\label{disc}

Exponential potentials in quintessence cosmologies have been recently
re-considered and offer interesting perspectives \cite{ru+sc02}. As an
example, an exponential potential can entail acceptable values for the
CDM density at the nucleosynthesis epoch.

A suitable choice of the exponent allows, through a transformation of
variables suggested by the N\"{o}ether symmetry approach \cite{cap+al96},
an exact integration of the Friedmann equations. This is the main
attractive feature of the model we have been considering in this
paper.

Scenarios alternative to a cosmological constant have been already
tested with gravitational lensing statistics. Flat universes with dark
energy with a constant equation of state has been considered
\cite{wa+mi98,cha+al02}. Such an ad hoc modelization of the dark
energy helps to discriminate between a cosmological constant and an
evolving quintessence but it does not account for a general treatment
of the dark energy component. On the other hand, physically motivated
scalar fields deserve particular attention. The model described in
Section~\ref{mode} presents, in my opinion, some very interesting
properties. In general relativity, it is quite exceptional to deal
with exact general solutions. This enables a comprehensive analysis
and a full treatment of the model and of its phenomenological
properties.

Apart from this general argument in favour of exact solutions in
general relativity, the examined exponential potential presents a
further appeal from the statistical point of view. Since it has only
two free parameters, i.e $\1$ and $H_0$, following Bayesian arguments,
it has the same a priori probability of a flat universe with a
cosmological constant and must be preferred to a flat quintessence
cosmology with a constant, but yet undetermined, equation of state of
the dark energy, $w_X$\footnote{These models are characterized by
three parameters: $H_0$, $\1$ and $w_X$.}. Despite of its simplicity,
the examined exact solution reflects many properties of the observed
universe, with the same statistical confidence of a constant
$\Lambda$-term \cite{did+al02,pav+al02,ru+se02}. It is compatible with
the supernovae data, whose analysis gives $\1 =0.15^{+0.15}_{-0.03}$
at 68\% confidence level \cite{pav+al02}. The peculiar velocity field
and the perturbation growth of structure can also be accounted for in
this scenario \cite{ru+se02}; data from galaxy clustering suggests a
value of $\1 = 0.18 \pm 0.05$ at 68\% confidence level. The considered
model, not including radiation, is realistic only in the contemporary
or, at most, the recent past regimes of the life of the universe.
However, a preliminary analysis of the anisotropies of the cosmic
microwave background also gives concordant results \cite{did+al02}.

In this paper, I have considered complementary constraints from
gravitational lensing statistics. This analysis is based on different
physics phenomena and on independent observations and can strengthen
conclusions on a cosmological model. An analysis of the CLASS data
sample provides a best-fit estimate, at the $68.3\%$ confidence level,
of $\1 = 0.31_{-0.14}^{+0.12}$ in the case of 11 lenses. When 10
lenses are considered, the value of the best estimate increases of
$\Delta \1 \simeq 0.04$. In addition to radio-selected sources,
optical surveys of quasars have been employed to constrain $\1$. This
analysis slightly prefers higher values of the pressureless matter
density, $\1=0.6 {\pm}0.2$. The simple fact that the results from the two
independent samples are compatible at the 1-$\sigma$ level is
encouraging and we can consider constraints on cosmological parameters
from gravitational lens statistics as quite reliable. Furthermore,
estimates from lensing statistics are also compatible with results
from both supernovae and galaxy clustering data.

A number of systematic effects can plague lens statistics analyses
\cite{cha03}. Whereas most of the possible sources of errors are well
controlled, the main uncertainty is connected with the adoption of the
galaxy LF. As seen in Sec.~\ref{resu}, the main contribution to the
error in the estimate of $\1$ derives from the uncertainties in the
early-type galaxy LF. Furthermore, there is no consensus among
observationally derived results for the early-type LF
\cite{lov+al92,cha03}. Actually, a reliable determination of the LF is
needed to firmly draw conclusions on cosmological parameters.
Nevertheless, as noted in Chae~\shortcite{cha03}, whenever the total
galaxy LF is reliably determined, conclusions on cosmological
parameters are accurate if the partition of the total LF into the
specific LF is careful.

\section*{Acknowledgements}
I thank C. Rubano and P. Scudellaro for the stimulating discussions on
the exponential potential. I also thank an anonymous referee for the
constructive comments.

\end{document}